# Combined stellar structure and atmosphere models for massive stars:

## Wolf–Rayet models with spherically outflowing envelopes

**Daniel Schaerer**

Geneva Observatory, CH-1290 Sauverny, Switzerland; e–mail: schaerer@scsun.unige.ch

**Abstract.** We present a simple analytical method to describe the structure of a spherically expanding envelope with strong mass outflow. The structure is consistently connected to the hydrostatic stellar interior and provides an adequate description of the outer boundary conditions for stellar models with large mass loss rates.

We apply our treatment to evolutionary models of Wolf–Rayet (WR) stars in order to study the possible influence of the stellar winds on the interior, and to determine more reliable radii of WR stars. Independently of the wind parameters (wind density, opacity, velocity law) the interior structure and evolution of WR stars is found to be unaffected by the outer layers.

On the other hand, the stellar parameters (radii, effective temperatures) may well depend on the wind structure. For hydrogen rich WR stars (WNL) we find the existence of a temperature domain in the HR–diagram, where a transient concentration of stars on their blueward track is predicted in case of a strong backwarming from the wind.

For WNE and WC/WO stars with strong mass loss rates we also derive subphotospheric radii corresponding to Rosseland optical depths of $\tau \sim 10\text{--}20$. The dependence of the subphotospheric radii on the adopted envelope structure is discussed. With respect to wind–free stellar models the subphotospheric radii are increased by up to a factor of $\sim 4$ for the most luminous WNE or WC stars. These radii and the corresponding effective temperatures should roughly be comparable with the stellar parameters ("core" radii and temperatures) of non–LTE atmosphere models of WR stars. Comparisons using the newly derived subphotospheric radii yields a better agreement with observations.

The stellar parameters obtained with the new treatment allow a better assignment of theoretical spectra to evolutionary tracks of evolved WR stars (WNE, WC). This also provides the base for future studies of the spectral evolution of post main-sequence massive stars and their descendants.

We also point out the possible importance of the iron opacity peak for the acceleration of WR winds in the optically thick part, which may be essential for the understanding of the dynamics of WR winds.

## 1. Introduction

Wolf–Rayet (WR) stars are among the stellar objects showing the largest mass loss rates observed ($\dot{M} \lesssim 10^{-4} \, M_\odot \text{yr}^{-1}$). As a consequence several complications arise for the derivation of stellar parameters and comparisons with evolutionary models, which mainly limit quantitative comparisons with observations to luminosities and surface abundances. The largest difficulty is the determination of a stellar radius, which is not unique (cf. Baschek et al. 1991). Essentially there are three reasons: *i):* As the photospheres of WR stars can be formed in the supersonic part of the wind, the density scale height is large, and the radius at a given optical depth becomes wavelength dependent (see e.g. Hillier 1987). *ii):* The density structure of the winds is not well known. Although it is widely thought that the basic driving mechanism of WR winds is radiation pressure (cf. Lucy & Abbott 1993), the dynamics of these stars is not yet understood (Schmutz 1994, 1995 and references therein). Therefore, the density structures are uncertain and have to be assumed or determined empirically. *iii):* With present state-of-the-art models we are still not able to account consistently for all relevant opacity sources in the expanding envelope (Schaerer et al. 1995c).

is also of interest in a more general context. Indeed wind–free evolutionary models predict an important population of WR stars with temperatures $\gtrsim$ 125000 K. If theoretical spectra from plane parallel atmosphere models corresponding to these temperatures are however assigned to this population, very strong nebular excitation is predicted, in contrast to e.g. observed H II regions (cf. García-Vargas & Díaz 1994). It is, however, also not correct to use photospheric temperatures ($T_{2/3}$) to characterise the spectral energy distribution (see below). Therefore, to obtain an adequate description of the emergent UV flux of a young stellar population including WNE and WC/WO stars requires a good knowledge of the fundamental parameters of these stars. This is also fundamental importance to determine the rôle of starbursts in AGN's (cf. Cid Fernandes et al. 1992, Leitherer et al. 1992).

Evolutionary models have so far treated stellar winds by, what I here call "*a posteriori*" methods. This means that the stellar interior (hydrostatic or hydrodynamic) is usually treated with a plane parallel grey atmosphere as the outer boundary condition. The spherical extension of the atmosphere is neglected, and no backscattering from the dense wind (also called wind blanketing) to the interior is taken into account – in other words the interior is completely unaware of the presence of the wind ! For photospheres of OB stars the effect of wind blanketing, which leads to backwarming and also modifies the spectral energy distribution, has been studied by Abbott & Hummer (1985). Standard evolutionary models account for the extension of the atmosphere by translating the stellar radius and the effective temperature $T_{\rm eff}$ of the wind–free star, in an independent "a posteriori" step, to a photospheric radius and corresponding effective temperature (called $R_{2/3}$, and $T_{2/3}$ respectively), which refer to an optical depth of 2/3.

Several approaches to obtain an "*a posteriori*" estimate of $R_{2/3}$ and $T_{2/3}$ have been made, taking either electron scattering opacity into account (de Loore et al. 1982, Langer 1989a), or including additional line opacity (Schaller et al. 1992), which yields a better agreement with observations (Maeder & Meynet 1994). On the other hand the spectral energy distribution of WR stars with identical photospheric temperatures $T_{2/3}$ can differ significantly. It is instead characterised by the so-called "core temperatures" $T_\star$ corresponding to approximately unit thermalisation depth, i.e. optical depths of 10–20 (Schmutz et al. 1992). A unique correspondence between a position in the HR–diagram and the stellar spectrum of WR stars thus requires the knowledge of "core" radii from evolutionary models. This is one aim of the present work.

The first combined stellar structure and atmosphere (*CoStar*) models, which consistently treat the stellar interior and their wind, have recently been presented by Schaerer (1995) and Schaerer et al. (1995ab, hereafter Pa-also predict the detailed spectral evolution for massive stars taking non–LTE effects and line blanketing into account. The atmosphere structure for OB stars consists of a quasi-hydrostatic spherically extended photosphere and the wind. One difficulty, however, arises for WR phases (e.g. WNE, WC) with very large wind densities, since sub-photospheric layers ($\tau \gtrsim 10$) may still be located at supersonic expansion velocities. To treat the zone reaching further down to the hydrostatic interior (subsonic velocities) with their detailed non–LTE radiation transfer calculations is clearly prohibitive. For WR stars an alternative treatment is thus required. Therefore, we here introduce a simple method to determine the temperature structure in a spherically expanding envelope, which may be of very large optical depth. The density- and velocity structure are parametrised and guarantee a smooth transition with the hydrostatic interior. Our formulation not only yields "core" radii characterising the emergent spectra, but it also allows to account consistently for the influence of the expanding envelope (backwarming) on the stellar interior. Due to the simplicity of the approach presented here, it can easily be applied to different WR phases, and the small parameter space is rapidly explored.

The outline of the present publication is as follows. First we summarise the physical ingredients of our models and develop the equations describing the expanding envelope (Sect. 2). Envelope models are presented in Sect. 3, where the influence of the modified boundary conditions is studied in detail. Consistent evolutionary models for WN and WC stars including the mass outflow are given in Sects. 4 to 6, where we also discuss its impact on the stellar parameters and the interior evolution. A short comparison with observations is made in Sect. 7. Finally the main conclusions are summarised in Sect. 8.

## 2. Input physics

The present calculations are based on the Geneva stellar evolution code. Taken apart the new treatment of the atmosphere and envelope described below, we use the same physical ingredients as the latest grids of Meynet et al. (1994). For a more detailed discussion of most ingredients the reader is referred to Schaller et al. (1992). In particular the models are computed with a moderate core overshooting of $d/H_P = 0.20$, where $d$ is the overshooting distance, and $H_P$ the pressure scale height at the boundary of the classical Schwarzschild core. For possible external convection zones we follow the treatment of Maeder (1987), which includes turbulent pressure, acoustic flux, and a density scale height (cf. Sect. 3). Radiative opacities are from OPAL (Iglesias et al. 1992), completed with the atomic and molecular opacities of Kurucz (1991) at low temperatures. For WR phases with modified carbon and oxygen surface abundances (WC), we use the recent

## 2.1. Spherically outflowing envelope and atmosphere

In order to consistently model the stellar wind and the interior we introduce a simple analytic method to treat the outflowing envelope and atmosphere. This treatment solves for the mass conservation, and radiative equilibrium in a spherically expanding medium. The momentum equation, which determines the velocity and density structure, is replaced by adopting a parametrised velocity law

$$v(r) = v_\infty \left(1 - \frac{R_0}{r}\right)^\beta, \qquad (1)$$

where $v_\infty$ is the terminal velocity. The free parameter $\beta$, with typical values of 1–2, expresses the steepness of the velocity field. $R_0$ will be determined below. Equation 1 and the continuity equation

$$\dot{M} = 4\pi r^2 \rho(r) v(r) \qquad (2)$$

determine the structure of what is henceforth called the expanding envelope. Mostly for convenience we chose the more general term "envelope" rather than atmosphere, since the total optical depth of this region can be substantially larger than what is usually understood as the atmosphere.

In order to allow a smooth transition to the hydrostatic interior model, the lower boundary of the expanding envelope is defined at a given velocity, which, as a simple working hypothesis, we choose at a constant fraction $s$ of the isothermal sound speed $a_\star$ of the perfect gas. Variations of $s$ also allow to explore the effect other definitions of the sound speed (e.g. adiabatic) might have. The inner boundary is thus given by

$$v_\star \equiv v(R_\star) = s\, a_\star, \quad \text{where} \quad a_\star = \sqrt{\frac{kT_\star}{\mu m_H}}. \qquad (3)$$

$\mu$ is the mean molecular weight, and $R_\star$ and $T_\star$ are the radius and effective temperature of the hydrostatic star respectively. From the continuity equation (Eq. 2) this condition determines the density $\rho_\star$ at the inner boundary, as well as $R_0 = R_\star \left[1 - (v_\star/v_\infty)^{1/\beta}\right]$ required in Eq. 1.

The temperature $T(R_\star)$ is calculated from the analytical solution of the spherically extended grey atmosphere in radiative equilibrium using the generalised Eddington approximation derived by Lucy (1971; cf. also Lucy & Abbott 1993). The temperature structure is given by

$$T^4 = \frac{3}{4} T_\star^4 \left(\tilde\tau + \frac{4}{3} W\right), \qquad (4)$$

where $W$ is the geometrical dilution factor defined by some photospheric radius $R_p \geq R_\star$. Note that the optical depth (Rogers 1993).

low). To clarify the non-unique definition of an "effective temperature" $T_{\rm eff}$ in objects with spherical extension, we repeat that it is defined from the total emergent luminosity $L = 4\pi r^2 \sigma T_{\rm eff}^4$. Thus, the effective temperature $T_\star$ of the hydrostatic star is e.g. given by setting $r = R_\star$ and $T_\star = T_{\rm eff}$. In the same way the effective temperature can be defined at any depth $r$. In particular $T_{2/3}, T_{10}, \ldots$ denote henceforth the effective temperature at given optical depths $\tau = 2/3,\ 10\ \ldots$

With the adopted velocity law, the effective optical depth $\tilde\tau = \int_r^\infty \kappa \rho \left(\frac{R_\star}{r}\right)^2 dr$ can be integrated analytically for a depth-independent opacity. For values of $\beta \neq 1, 2$, and 3 this yields

$$\tilde\tau(x) = C\, [(\beta-1)(\beta-2)(3-\beta)]^{-1} * \\ [\{x^3(\beta-1)(\beta-2) + x^2\beta(1-\beta) + 2(x\beta-1)\} \\ (1-x)^{-\beta} + 2] \qquad (5)$$

where

$$C \equiv \frac{\kappa \dot{M}}{4\pi v_\infty R_\star}, \qquad (6)$$

and $x = R_0/r$. Note, however, that the optical depth $\tau = \int_r^\infty \kappa \rho dr$ is given by

$$\tau(x) = C(1-\beta)^{-1}\left[1 - (1-x)^{1-\beta}\right] \quad \text{for } \beta \neq 1. \qquad (7)$$

Unless specified otherwise, the opacity $\kappa$ is taken as the electron scattering opacity $\sigma_e$ of a fully ionised mixture with abundances from the stellar interior model at the given stage.

Since we are dealing with optically thick envelopes one has $W = 1/2$ at the inner boundary $R_\star$, and the temperature $T(R_\star)$ is thus solely determined by $\tilde\tau$. The variable $C$ describes the behaviour of $\tilde\tau$ on the basic wind parameters ($\dot{M}$, $v_\infty$, opacity, and radius $R_\star$). From this we immediately see that for a given velocity law the temperature at the inner boundary $R_\star$ increases with mass loss, increasing opacity, decreasing terminal velocity, and decreasing stellar radius, as expected.

The dependence of the temperature on the shape of the velocity law is illustrated in Fig. 1 for two different values of $v_\star/v_\infty$. The solid line ($v_\star/v_\infty = 0.01$) represents a typical value for WN stars, if we set $v_\star$ equal to the sound speed $a_\star$. For comparison, a case where the inner boundary is defined at lower velocity is also shown (dashed). Figure 1 shows that for all values of $\beta \gtrsim 1$ the depth scale $\tilde\tau(R_\star)$, and hence the temperature at the inner boundary of the wind does not strongly depend on the shape of the velocity law. For $v_\star/v_\infty = 0.01$ e.g., the temperature at the bottom of the wind varies by less than $\sim 15$ %. The changes resulting from the adopted value of $v_\star$, and the uncertainties in the opacity $\kappa$ are clearly larger than the exponent of the velocity law (cf. Sect. 3).

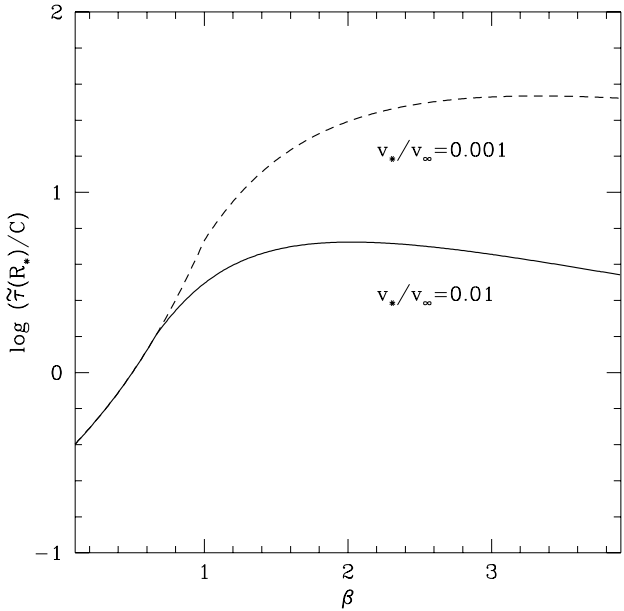

**Fig. 1.** Logarithm of $\tilde{\tau}/C$ at the inner boundary $R_\star$ as a function of the exponent $\beta$ of the velocity law (Eq. 1). $R_0$ is implicitly defined by the choice of $v_\star/v_\infty$. The solid line correspond to the typical value of $v_\star = a_\star$ for WN stars, where $a_\star \sim 20$ km s$^{-1}$, and $v_\infty \sim 1000$–$2000$ km s$^{-1}$. Note that for all values of $\beta \geq 1$ the depth $\tilde{\tau}(R_\star)$ varies by less than $\sim 0.23$ dex, which corresponds to temperature changes of less than $\sim 15\%$

Once the temperature $T(R_\star)$ is determined ($\rho_\star$ is also known), we finally compute the total pressure at $R_\star$, which is also required for the integration of the stellar structure equations. The pressure is derived from the equation of state taking into account partial ionisation of H, He, C, O, Ne, and Mg (cf. Schaller et al. 1992). It is straightforward to replace the usual boundary conditions of stellar interior models (plane parallel grey atmosphere, cf. Kippenhahn & Weigert 1990) with the method presented above. The only additional parameters, which need to be specified, are the wind velocity law ($v_\infty$, $\beta$) and the wind opacity. The terminal velocity $v_\infty$ can be obtained from observations. The mass loss rate has already been taken into account to determine the mass removed at each evolutionary time step.

With this procedure we have thus obtained an adequate method to describe the outflowing envelope and atmosphere with few basic parameters, which describe the hydrodynamic structure from close to sonic velocities up to the highly supersonic asymptotic wind velocity. Assuming the velocity structure is monotonous in the entire star, the choice of the velocity at the inner boundary guarantees that below hydrostatic equilibrium should be closely verified, which means that hydrostatic interior solutions are clearly appropriate.

Before turning to the remaining physical ingredients of our models, it is worth clarifying some problems relating to the radius and temperature definition of the extended atmospheres of WR stars (see also Hillier 1991, Schmutz et al. 1992, Schmutz 1995). As already mentioned the major difficulties stem from the hydrodynamic structure (especially at large optical depths) and the wind opacity, which are badly known. For the first point we have adopted a parametrised description as a working hypothesis (see Sect. 2.1). A large uncertainty in our treatment arises from the difficulty of specifying the wind opacity. In this context it is, however, important to clearly distinguish the two following points: *(i)* the cooling and heating processes, which determine the temperature structure (including the backwarming), and *(ii)* the (frequency dependent) continuum opacity allowing to define the radius, where radiation is thermalised. Let us therefore discuss these points separately.

*(i)* As will be shown in Sect. 4, the effect of backwarming can have an influence on the outer layers of WN stars with hydrogen, and lead to a relative increase of the stellar radius. The large opacity adopted in Sect. 4 (20 times the electron scattering opacity), corresponds to the typical value proposed by Lucy & Abbott (1993), which is derived from the effective scattering coefficient ($\sigma_{\rm eff}$) required to explain the dynamical wind structure by radiation pressure. $\sigma_{\rm eff}$ is thus a flux weighted mean opacity, which shows that the Lucy & Abbott approach for the determination of the temperature structure is basically equivalent to the treatment of Schaller et al. (1992) to include the effect of line blanketing. Whether, however, the procedure using flux weighted mean opacities indeed yields a good description of the detailed heating and cooling mechanisms including line blanketing is not clear. The present method of Schaerer & Schmutz (1994) applied to WR stars gives only a modest temperature increase (cf. Schaerer et al. 1995c). This question awaits a quantitative treatment including the calculation of thermal and statistical equilibrium.

*(ii)* Given a strong variation of the continuum opacity with wavelength, any radius definition in extended atmospheres is necessarily wavelength dependent (e.g. illustration in Hillier 1987, Fig. 2). For WNE stars e.g., the electron scattering depth $\tau_e = 2/3$ yields a good estimate of the thermalisation optical depth $\tau_{\rm th} = 2/3$ at UV to visible wavelengths (see Schmutz 1995), and hence for the photospheric radius characterising the largest fraction of the directly observable flux. On the other hand, to obtain a radius completely independent of frequency requires radiation at *all wavelengths* to be thermalised, which is typically the case at optical depths $\tau_{\rm Ross} \sim 10$–$20$. Referring to these optical depths therefore allows the physically most relevant trum including the EUV flux shortward of the Lyman edge.

For the non–LTE conditions in WR winds the Rosseland opacity is mostly dominated by electron scattering. Therefore the optical depths, and hence the radii $R(\tau \gtrsim 10)$ predicted in this work, should be similar to those from non–LTE model atmospheres, provided the other parameters are identical. Changes of the opacity due to the ionisation stratification, are not taken into account in our procedure, but should be of minor importance.

To complete the physical ingredients of our models we shall now turn to the description of the wind parameters.

### 2.3. Mass loss rates

Since mass loss is of particular importance for WR phases, the adopted prescriptions shall briefly be recalled here. Remember that in our evolutionary models the entry in the WR phase is defined by a hydrogen surface abundance $X < 0.4$ in mass fraction, and $T_{\rm eff} > 10000$ K.

- For the WNL phase we take an average mass loss rate of $8\,10^{-5}\,M_\odot\,{\rm yr}^{-1}$, as in Meynet et al. (1994). Note that the observational scatter is found to be $\sim 0.4$ dex, with a mean being slightly lower than the adopted value (Hamann 1994, Crowther et al. 1995b).
- For WNE and WC/WO stars, we use the mass dependent mass loss rates from Langer (1989b), as in previous evolutionary grid calculations (Schaller et al. 1992, Meynet et al. 1994). The adopted mass loss rates show a good agreement with the observations presently available (cf. Hamann 1994).

### 2.4. Wind velocities

In this paragraph we describe the choice of the adopted terminal velocities for WR stars.

#### 2.4.1. WN stars

Observed terminal velocities of WN stars are in the range of $\sim 700$ to 2500 km s$^{-1}$. Earlier subtypes tend to have higher $v_\infty$ although the relation shows a large scatter (e.g. Hamann et al. 1993, Crowther et al. 1995b). For WR stars no clear picture has emerged yet, which would relate the terminal velocity to fundamental stellar parameters. In analogy with OB stars, one can suspect a correlation of $v_\infty$ with the escape velocity. Using the stellar parameters of WNL stars from evolutionary models (cf. Meynet et al. 1994) and applying the same proportionality as observed for O stars ($v_\infty = 2.78\,v_{\rm esc}$, Groenewegen et al. 1989), this yields in fact values, which are compatible with the observations. On the other hand, one is obviously tempted to relate the trend of higher $v_\infty$ for crease of $v_\infty$ towards larger temperatures, which is found in the data of Hamann et al. (1993), can be fitted with the following formula:

$$v_\infty = (2395.87 \pm 244.83) \log T_\star - (9656.98 \pm 1125.04), \quad (8)$$

where the residual rms is 241.61 km s$^{-1}$. At temperatures $\log T_\star < 4.24$ we adopt $v_\infty = 500$ km s$^{-1}$, which corresponds to typical values for Ofpe/WN9 stars (e.g. Crowther et al. 1995a). For the purpose of this work, Eq. 8 is clearly sufficient to reproduce the observed trend. For most calculations we shall however simply use a constant value of $v_\infty = 1400$ km s$^{-1}$, which is close to the mean value of 1347.5 km s$^{-1}$from the data of Hamann et al. (1993). As will become clear later (see Sect. 4), the exact choice of the terminal velocity in the WN phases is, however, not crucial for our results.

#### 2.4.2. WC stars

For Wolf-Rayet stars from the WC/WO sequence, a clear correlation of the terminal velocity with spectral class is well established (cf. Abbott & Conti 1987) and has been been confirmed by recent studies, which take turbulence into account to obtain accurate line profile fits (Eenens & Williams 1994). On the other hand there is also ample evidence for a correlation of the C/He abundance ratio with WC/WO subtype (Smith & Hummer 1988, Eenens & Williams 1992) and possibly also for O/He (Nugis 1991). Adopting the relation between (C+O)/He and the WC/WO subtype from Smith & Maeder (1991), one is thus able to assing a wind velocity provided the abundances of He, C, and O are known.

The method is the following: We adopt the terminal velocities from Eenens & Williams (1994) derived from IR lines taking turbulence into account. The observations include 18 WC stars, and cover the largest subtype range for which homogeneous data is available. Their values are plotted in Fig. 2 (triangles). They can well be fitted by

$$v_\infty = (3738.49 \pm 236.18) - (281.40 \pm 31.07){\rm WC}, \quad (9)$$

with a residual rms of 190.85 km s$^{-1}$. WC stands for the subtype number, and $v_\infty$ is to be taken in km s$^{-1}$. As a comparison we have also plotted the values derived from UV lines by Prinja et al. (1990, squares) together with a fit to their data, which show a slightly steeper trend. If the latter relation is extrapolated to WC4, we see that the agreement between both fit relations is of the order of $\lesssim 10$ % for $v_\infty$, which is clearly sufficient for our purposes. The terminal velocities for WC stars with strong lines (WC-s, Koesterke & Hamann 1994) agree well with the above relation, while their weak lined WC stars (WC-w) have considerably larger wind velocities.

From Smith & Maeder (1991) we now take the mean values of (C+O)/He (in number) to be representative for the corresponding WC subclass, i.e. WC9: 0.045, WC8: 0.1, WC7: 0.2, WC6: 0.3, and WC5: 0.55. A linear least square fit yields $(C+0)/He = 1.086 - 0.121\,WC$, where again WC is the subtype number. Combined with Eq. 9 we can express the terminal velocity as a function of $(C+O)/He$, which yields (in km s$^{-1}$)

$$v_\infty\left(\frac{C+O}{He}\right) = 3738.49 - 2324.36\left(1.086 - \frac{C+O}{He}\right). \quad (10)$$

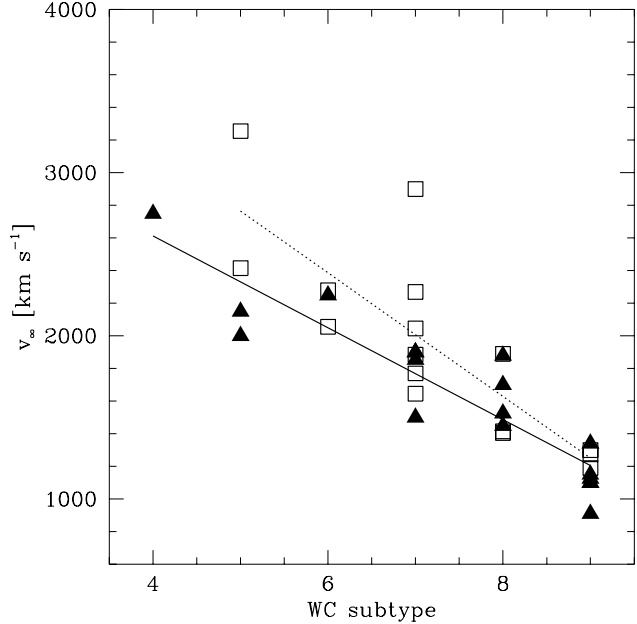

**Fig. 2.** Observed terminal velocities for WC stars as a function of subtype. Triangles: 18 stars from Eenens & Williams (1994), open squares: 16 measurements from Prinja et al. (1990). The solid line shows the fit of Eq. 9 to the values of the former authors, the dotted line a fit to the Prinja et al. data

This formula will be used to determine $v_\infty$ in WC/WO phases. Despite its simple derivation it reproduces well the observed trend of increasing wind velocities for WC stars, and also yields a good agreement with observed $v_\infty$ values of WO stars (Kingsburgh & Barlow 1991), when Eq. 10 is extrapolated towards larger $(C+O)/He$ values than formally allowed.

## 3. Envelope models – the influence of the mass outflow

In this paragraph we construct envelope models including the spherically expanding stellar wind and the lower hydrostatic layers. This allows us to study the influence of the backwarming effect of the wind on the external layers of the stellar structure. An envelope model describes the outermost stellar layers and is defined by the outer boundary conditions (cf. Table 1), the atmospheric boundary conditions (Sect. 2.1), and the mass it embraces (here region treated here covers only a very small percentage of the total mass, its extension corresponds to $\sim 60 - 90\%$ of the outer radius. We solve for the continuity equation, hydrostatic equilibrium, and the energy transport allowing both for radiative or convective transfer. In a next step (Sects. 4 and 6) we will then present evolutionary tracks, where the entire stellar structure is solved consistently.

### 3.1. Calculated models

To illustrate the envelope structure and its dependence on different boundary conditions we have chosen a model corresponding to the WNL phase from the 85 $M_\odot$ track calculated with our "combined stellar structure and atmosphere models" (Schaerer 1995, Schaerer et al. 1995a). The stellar parameters are summarised in Table 1. The temperature corresponds to a domain, where WNL stars are found to be sensitive to boundary conditions ("WNL strip" see Sect. 4). Qualitatively the results do not depend on the adopted luminosity. Therefore the discussion in this paragraph also applies to WN models of lower initial masses located in the "WNL strip".

**Table 1.** Adopted stellar parameters for envelope models

| | |
|---|---|
| $T_\star$ [K] | 38700. |
| $R_\star/R_\odot$ | 25. |
| $\log L/L_\odot$ | 6.10 |
| $\dot{M}$ [$M_\odot\,yr^{-1}$] | $8.\,10^{-5}$ |
| $v_\infty$ [km s$^{-1}$] | 1500. |
| $M/M_\odot$ | 39.1 |
| $X$ (mass) | 0.08 |
| $Y$ (mass) | 0.90 |

The influence of the atmospheric boundary condition on the envelope solutions can be illustrated with several models, which are summarised in Table 2. The first model has been computed with a plane parallel grey atmosphere in the Eddington approximation. The inner boundary of the atmosphere is set at $\tau_{Ross} = 2/3$. The remaining models (2–4) have been calculated with the outflowing atmosphere models from Sect. 2.1. Their inner boundary is defined by the velocity $v_\star$, for which we take values no larger than the isothermal sound speed $a_\star$ to guarantee a smooth transition with the hydrostatic part. Models 2 and 3 explore the influence of the opacity in the wind. The value for model 2 corresponds to the electron scattering $\sigma_e$ for the entirely ionised mixture. In model 3 we enhance the scattering opacity to simulate strong line blanketing. The adopted enhancement factor of 20 is consistent with the effective scattering coefficient required if radiative driving dominates the dynamics of WR winds (cf. Lucy & Abbott 1993). Finally, in model 4 the inner boundary is modi-fied by setting $v_\star = 0.1\,a_\star$, otherwise keeping the same parameters as model 3.

**Table 2.** Summary of envelope models. P indicates plane parallel atmosphere, W wind. The input parameters are $v_\star$, $\beta$, and the wind opacity $\kappa$ (cf. 2). The other quantities denote the value of the optical depth, temperature, pressure, and density at the inner boundary of the atmosphere

| model | P/W | $v_\star$ [km s$^{-1}$] | $\beta$ | $\kappa$ [cm$^2$ g$^{-1}$] | $\tau$ | $\log T$ | $\log P$ | $\log \rho$ |
|---|---|---|---|---|---|---|---|---|
| 1 | P | | | | 2/3 | 4.59 | 3.76 | −10.2 |
| 2 | W | 16.4 | 2.1 | 0.21 | 3.2 | 4.65 | 4.03 | −10.1 |
| 3 | W | 16.4 | 2.1 | 4.21 | 63.8 | 4.95 | 5.19 | −10.1 |
| 4 | W | 1.64 | 2.1 | 4.21 | 210.2 | 5.11 | 5.85 | −9.1 |

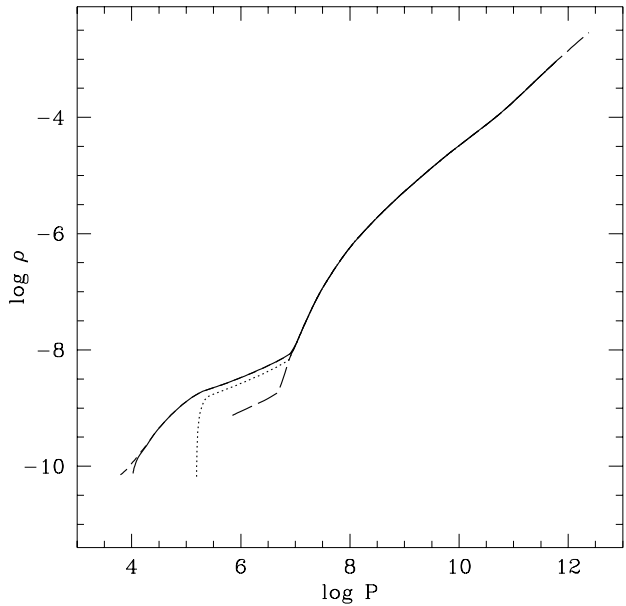

**Fig. 3.** Density structure of the envelope models 1 to 4, plotted as a function of $\log P$ used as the depth variable. The symbols are as follows: model 1 (short-dashed), 2 (solid), 3 (dotted), and 4 (long-dashed). Units on all structure plots (Figs. 3 to 7) are in cgs

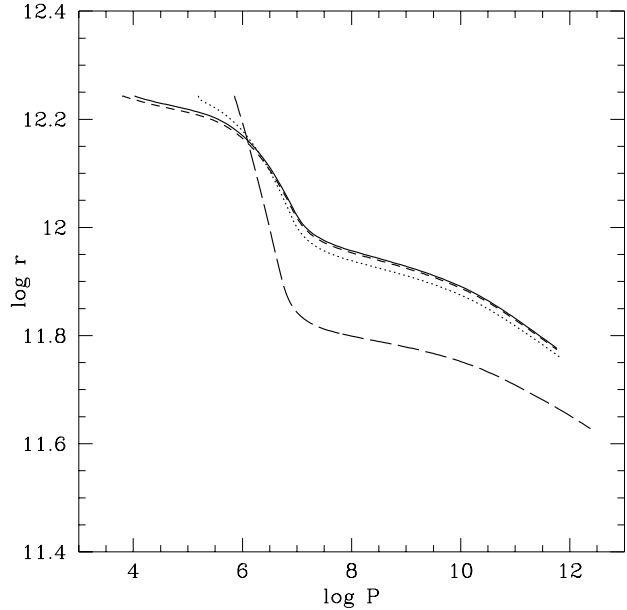

**Fig. 4.** Radial extension of the envelope models (same symbols as in Fig. 3). All models have a given external radius (cf. Table 1). The inner boundary is set at $M_r/M = 0.997$. Note in particular the difference of the $r$ gradient in outer parts of the envelope, which leads to the different inward extensions

### 3.2. Discussion

In Figures 3 and 4 we have plotted the run of the density and radius of all models. The total pressure serves as a depth variable[1]. The model with the plane parallel atmosphere providing the outer boundary conditions (model 1), and the wind model 2 show essentially the same structure. Despite the small temperature increase in model 2 (cf. Table 2) they are nearly indistinguishable in these figures. For the parameters of model 2 the wind effects are therefore negligible.

Both models 3 and 4 have a higher pressure at the outer boundary due to the strong "backwarming" from the wind. The case differing most from the plane parallel atmosphere is clearly model 4, since it also has a larger density. How does this influence the depth of the envelope? From Fig. 4 we see that the models show different inward radial extensions. This behaviour is simply under-

---
[1] Note that since radiation pressure strongly dominates, and the temperature gradient is essentially independent of $P$ and $T$ in the entire envelope, all temperature profiles would lie on the same line if plotted as a function of the pressure. They are therefore not shown here.

be written as

$$\frac{\partial \log r}{\partial \log P} = -\frac{rP}{GM_r\rho} \quad (11)$$

in Lagrangean coordinates. In the envelope $M_r$ is nearly constant, and the outer value for the radius is fixed ($r = R_\star$). Therefore both a higher pressure and lower density imply a steeper $r$ gradient, and lead to a larger extension of the envelope. Considering variations of the outer boundary conditions, a large envelope extension can thus be obtained by increasing the ratio of $P/\rho$. If the density at the outer boundary can be kept constant, a temperature increase (due to larger opacity, or a change of $\beta$ as illustrated in Fig. 1) at the bottom of the atmosphere thus leads to a larger inward extension. This explains the differences between models 1+2 and model 3, which are, however, rather small. The main reason for the small radius difference is that in model 3 the density rapidly converges to the solution of the "classical" model 1.

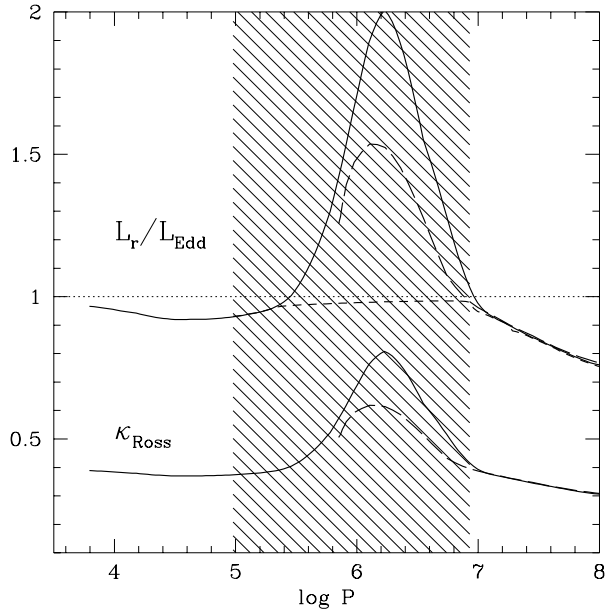

**Fig. 6.** Detailed view of the outer zones of models 1 and 4. The shaded area is as in Fig. 5. Shown are the Rosseland mean opacity (in cm$^2$ g$^{-1}$), and the total local luminosity $L_r$ in Eddington units for model 1 (solid), and 4 (long dashed). The short-dashed line shows that the *radiative* luminosity of model 1, using the density scale, never exceeds the Eddington luminosity. Discussion in text

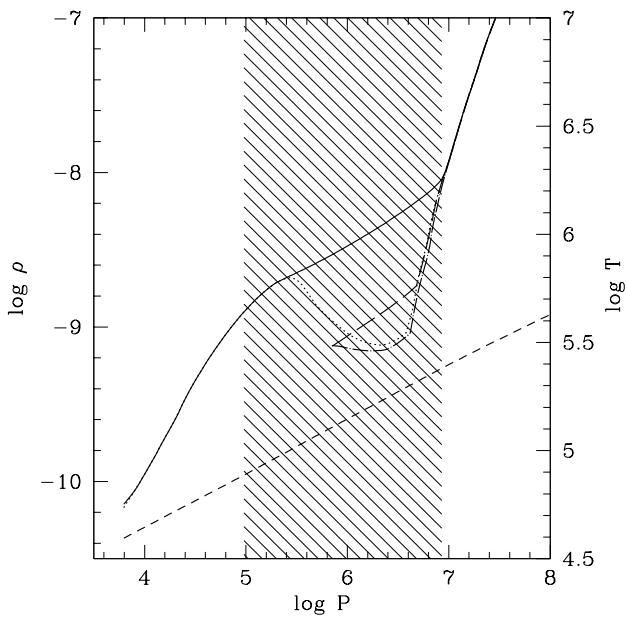

**Fig. 5.** Detailed view of the outer zones of models 1 (plane parallel atmosphere) and 4 (wind). The shaded area denotes the convective zone of model 1, which is due to the Fe opacity peak. In model 4 the convection zone starts at its outer boundary and has the same inward extension as model 1. Shown is the density structure as a function of the total pressure. Structure of model 1 (solid), and model 4 (long dashed) calculated with a density scale height and $l/H_\rho = 1$. Calculations adopting the pressure scale height yield density inversions: dotted line (model 1), dashed-dotted line (model 4). The short-dashed line shows the temperature (scale on the right)

To understand the important increase of the extension of model 4 it is worth recognising an additional "fortuitous" complication, which turns out to be relevant for determining stellar radii of hydrogen rich WR stars. We have therefore plotted an enhanced view of the outer envelope structure of the models with the largest differences (models 1 and 4) in Figs. 5 and 6. The point of particular importance is the finding of a small convective zone, which is present in all our models. The convection zone, also noted by Glatzel & Kiriakidis (1993), Alberts (1994), and Schaerer et al. (1995c), is due to the opacity peak of iron (cf. Iglesias et al. 1992), situated around $\log T \sim 5.2$ (see Fig. 6).

The key point to explain different extensions is that in the convective zone the density increases less steeply than in the radiative layers. The reason for this will be discussed below. If we consider e.g. model 4, we see that due to the convective layer, a lower density is maintained between $6 \lesssim \log P \lesssim 6.8$ compared to models 1–3. Therefore, according to Eq. 11, the radius of model 4 decreases more rapidly in this interval, which leads to the final $\sim 40$ % difference in radius.

Let us now examine the behaviour of the density structure within the convection zone. The approximately constant slope in the $\log \rho$–$\log P$ diagram (Fig. 5) is due to adopted treatment using a density scale height and $l/H_\rho = 1$ as in Maeder (1987). If instead we use the pressure scale height and $l/H_P = 1.6$, we obtain a density inversion, which occurs where the radiative acceleration exceeds the local gravity $g$, i.e. $\kappa L_{\rm rad}/(4\pi r^2 c) > g$. This is illustrated in Fig. 6 for models 1 and 4. We note that the finding of a density inversion is not due to a hydrostatic treatment. As mentioned above, for MS stars Glatzel & Kiriakidis (1993), using a hydrodynamic code, also find a $\rho$–inversion related to the Fe opacity peak (cf. Gautschi & Glatzel 1990). This may, however, also be related to the use of time steps, which are larger than the relevant timescales (Meynet, 1994).

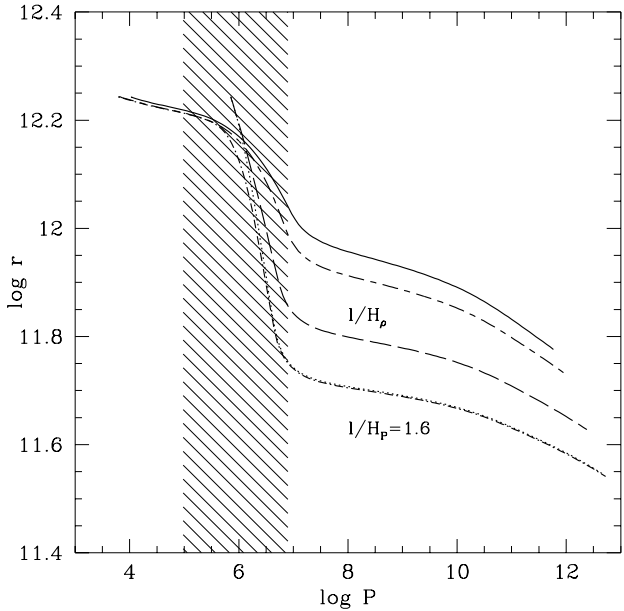

**Fig. 7.** Run of the radius as a function of the total pressure depending on different treatment of convection. The shaded area denotes the convective zone of model 1 (cf. Fig. 5). As reference models we plot model 1 (solid) and 4 (dashed) both computed with $l/H_\rho = 1$. The long-dashed–short-dashed line is model 1 but adopting $l/H_\rho = 0.5$. The models showing the largest extension are those showing the density inversion ($l/H_P = 1.6$ model 1 and 4, dotted and dashed-dotted lines, nearly coinciding)

Before we discuss its impact on the extension of the envelope, it is worth clarifying the question of density inversions in this context (see Maeder 1992 for a recent review). First, we note that the region with the $\rho$-inversion is not Rayleigh-Taylor unstable, since there the effective gravity is directed outwards (cf. Wentzel 1970). Similarly, the strong supra-Eddington luminosity observed within the convection zone (Fig. 6) is simply due to the fact that other modes of energy transport contribute with a significant fraction to the total energy transfer (up to ∼ 20 % convective, and 40 % acoustic flux respectively in the considered cases). If the density inversion is suppressed minosity obviously never exceeds the Eddington limit. In other words, the existence of the convective zone is due to the strong increase of the opacity, which, according to the treatment may or may not lead to a density inversion. The (total) supra-Eddington luminosity is related to the strong contribution of non-radiative fluxes, and does, however, *not necessarily* lead to an outward acceleration of matter, as suggested in particular by Kato & Iben (1992) for WR stars. Nevertheless, we note that in the outer layers the effective gravity is indeed strongly reduced, indicating that "potentially" there is a large radiative force available, which might contribute to initiate the mass outflow (cf. Sect.4).

Given the dependence of the total envelope extension on its outer density structure, and hence on the treatment of convection, we have also investigated its influence on models 1 and 4. In Figure 7 we have plotted the run of the radius using either the density scale height (and $l/H_\rho = 1$ or 0.5) or the pressure scale with $l/H_P = 1.6$. As expected, the models having the lowest density in the outer parts show the largest radial extension. Using the pressure scale height the wind model with strong backwarming (model 4) in fact turns out to be nearly identical to the plane parallel case. For the present model parameters, changes due to the treatment of convection and different atmospheres are thus of similar order.

Let me summarise. We have seen that for typical effective temperatures of WNL stars envelope solutions are sensitive to the treatment and changes of atmospheric boundary conditions. Changes are indeed expected between plane parallel atmospheres, which are certainly not adequate for stars with strong stellar winds, and a more consistent treatment including an expanding atmosphere with wind blanketing. The basic result is that the spherical extension of the envelopes may be increased. This finding is intimately related to the presence of a small layer with a considerable opacity increase (due to iron), which is unstable to convection. Depending on the effective temperature, i.e. the depth where this layer is located, its importance for determining the envelope extension and hence the stellar radius is expected to vary crucially. This will be studied in the following section, where we present evolutionary models covering the WN phase.

## 4. WN stars with expanding atmospheres

Let us now have a look at evolutionary tracks including spherically expanding atmospheres. We shall first discuss WNL stars, which are defined by the presence of both hydrogen and helium on their surface. The more evolved WNE and WC/WO phases are covered in Sects. 5 and 6.

To study the influence of the atmospheric treatment we have calculated two evolutionary tracks for initial masses

Note that all temperatures refer to the effective temperature $T_\star$ of the hydrostatic star (cf. Sect. 2.1). For the WNL phase considered in this paragraph, however, the extension of the layers between the hydrostatic interior and electron scattering optical depths of $\tau = 20$ or 10 is small. Therefore we have $T_\star \approx T_{20} \approx T_{10}$. The largest difference amounts to $\sim 0.1$ dex (see below).

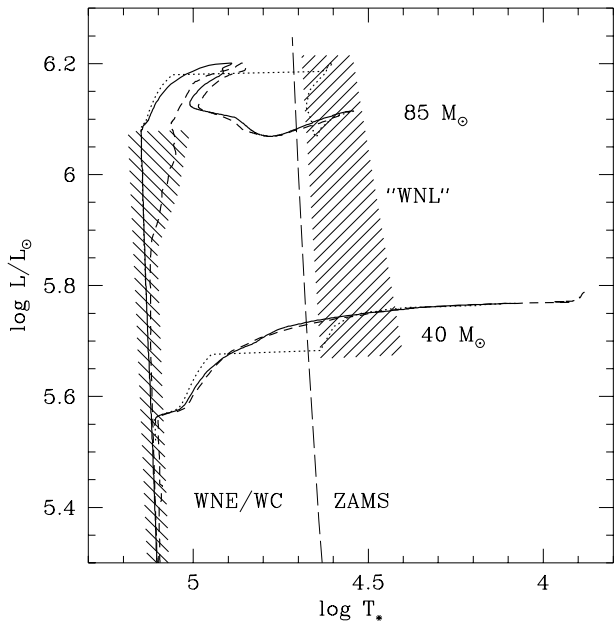

**Fig. 8.** HR diagram for WR phases of the 40 and 85 $M_\odot$ model. Remember evolution proceeds bluewards and to lower luminosities. Solid lines show tracks with plane parallel atmospheres, dashed lines with spherically expanding envelopes/atmospheres. For strong wind blanketing one obtains the dotted tracks. Shaded areas indicate the position of the WNE and WC phases at $\log T_\star \sim 5.1$, and the suggested strip for WNL stars predicted with strong blanketing (marked "WNL"). The ZAMS from Meynet et al. is also plotted (long-dashed)

In Figure 8 we have illustrated several tracks for each initial mass. First, the dashed line shows the results from standard models using the plane parallel atmosphere. Adopting the spherically expanding atmosphere with $v_\star = a_\star$, $\beta = 2.1$, $v_\infty$ from Sect. 2.4, and $\kappa = \sigma_e$, yields the solid line track. We immediately see that the differences are negligible for the WNL phases. This is basically due to the "moderate" wind density and low opacity, resulting in relatively small total optical depths. The largest values obtained are $\tau(R_\star) \sim 10$–30 for $\log T_\star = 5.$, which implies a low backwarming from the extended atmosphere. This is in line with the results of the envelope models 1 and 2 (Sect. 3).

---

[2] The pre-WR tracks correspond to Meynet et al. (1994) for the 40 $M_\odot$ model; the 85 $M_\odot$ track is from the "combined stellar structure and atmosphere models" of Schaerer (1995) and Schaerer et al. (1995a).

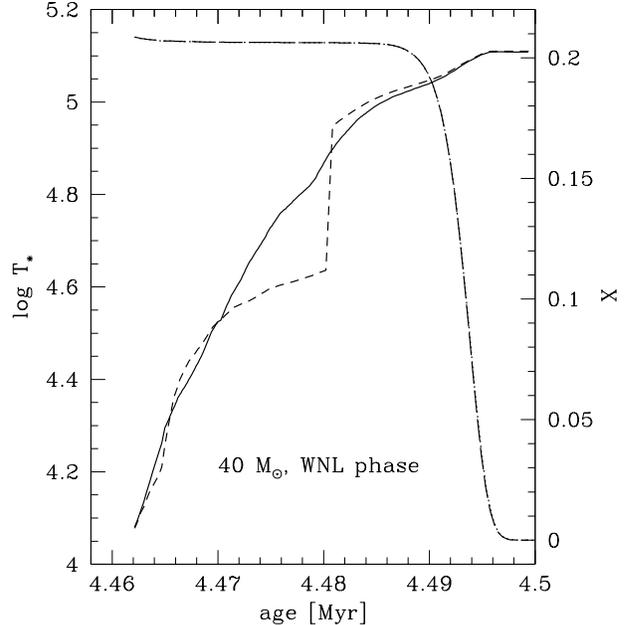

**Fig. 9.** Time evolution of the temperature $T_\star$ during the entire WNL phase of the initial 40 $M_\odot$ model. Tracks are plotted with the same symbols as in Fig. 8. The hydrogen surface abundance corresponding to both models is shown by the long-dashed and dotted lines (scale on the right)

On the other hand, if we adopt a larger wind opacity ($\kappa = 20\,\sigma_e$), the backwarming effect is considerably enhanced (cf. models 2, 3). The resulting evolutionary path is shown by the dotted line in Fig. 8. Overall the evolutionary paths, and in particular also the deep interior and its properties (lifetimes, internal abundance profiles etc.), remain unaffected. Nevertheless one very interesting feature can be observed from the tracks with strong wind blanketing. During the bluewards evolution in the WNL phase the models are temporarily stopped, or concentrated in rather small strip at temperatures $4.4 \lesssim \log T_\star \lesssim 4.65$, which has been marked in Fig. 8. The timely behaviour is illustrated in Fig. 9. While most models advance continuously towards the blue, the model with strong backwarming maintains a larger radius, thereby staying at $T_\star \lesssim 44000$ K, before it transits in a short lapse of time to larger temperatures. The same behaviour is also found for the 85 $M_\odot$ track, and is luminosity independent. It rests, however, on an strong backwarming effect, which is obtained by adopting a large opacity in our calculations. As expected (cf. Sect. 3), modifying the other parameters is found to be of less influence.

from the considerations of Sect. 3. In this temperature domain, the small convective zone related to the Fe opacity is sufficiently close to the surface so that the pressure and density is still affected by the external boundary conditions. In this case the buffering high opacity zone is capable to maintain a large radius, as explained in Sect. 3. The blueward trend, resulting from the homogeneisation due to mass loss, can however not be stopped. Once the Fe convection zone has moved sufficiently outwards, density and pressure profiles rapidly converge to the same structure, which is nearly independent of the outer boundary conditions. Thus the bluewards transition occurs.

As shown in Sect. 3, the importance of the convection zone for the envelope extension also depends on the treatment of the possible density inversion. Calculations using either the density- or the pressure scale height, however, yield the same "strip", and show the same dependence on the other parameters as discussed above. A brief comparison with observations of WNL stars in given in Sect. 7.1.

### 4.1. Radiation driven winds of WR stars – the rôle of Fe

We have seen that the strong opacity peak at $\log T \sim 5.2$, due to iron, may play an important rôle in determining the structure of the outermost layers and thereby the stellar parameters of WNL stars (cf. Sect. 3). In all cases the Fe opacity leads to a strong decrease of the local effective gravity. In our WNL models this domain is located in subsonic layers, whereas along the entire WNE and WC sequence (see below) the zone with the opacity peak is encountered in the expanding envelope, i.e. at $v > v_\star$. By specifying the envelope structure and adopting a constant opacity it has therefore completely been neglected. As will be shown below this has no implications on the interior structure and evolution. However, we strongly suspect that the large Fe opacity bump may be an important contributor to the acceleration of the stellar winds of hot WR stars. Similarly it may also be the most important blanketing agent in the deep photospheric layers (cf. Schaerer et al. 1995c). Observationally Fe v and vi is known to be responsible for strong blanketing effects in WNE stars (cf. Koenigsberger 1990), while higher ionisation stages will dominate at larger depths. A correct treatment of Fe in line blanketed atmosphere models is however not yet possible (Schaerer et al. 1995c). More quantitative work will be required to understand the optically thick winds of WR stars. Implications of the above suggestion (metallicity dependence !?) and the search for possible observational constraints also deserve futures studies.

## 5. WNE stars

Let us now study more evolved WR phases. In this paragraph and the following ones, we have adopted the following parameters to describe the expanding envelope: of these parameters will be discussed. First we shall concentrate on the radii of WNE stars. The HR diagram from the 40 and 85 $M_\odot$ tracks for the advanced WR phases is presented later (Fig. ??).

After the evolution through the hydrogen rich WNL phase, evolutionary models predict a phase showing an essentially pure He surface abundance. In this phase, identified with WNE, and the subsequent WC/WO phases, the basic parameters of WR stars follow several simple relations (cf. Schaerer & Maeder 1992). The fundamental link is the well known mass-luminosity relation (Vanbeveren & Packet 1979, Maeder 1983, Langer 1989a, Schaerer & Maeder 1992). In particular, a well defined mass-radius relation is obtained, independently of whether the stellar interior is assumed to be in hydrostatic equilibrium or a standard hydrodynamic treatment is accounted for (see Fig. 10).

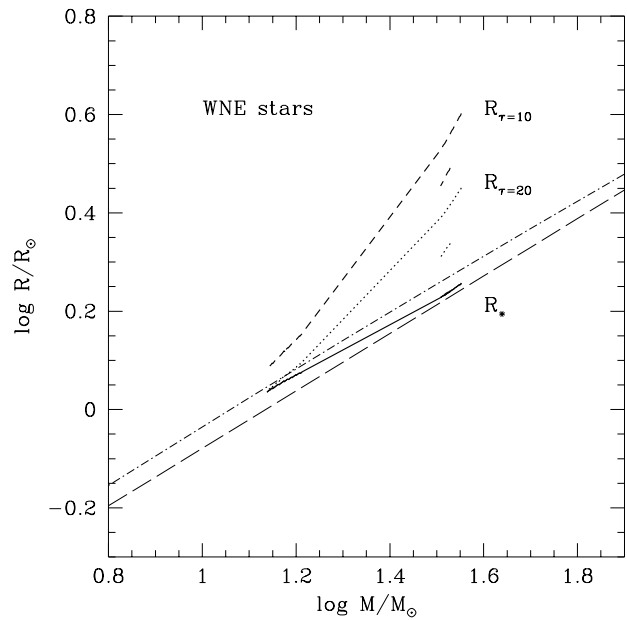

**Fig. 10.** Mass–radius relation for WNE stars. The solid line shows the radius $R_\star$ of the hydrostatic interior. The subphotospheric radii $R_{10}$ and $R_{20}$, adopting $\beta = 2.1$, are shown by the short-dashed and dotted line respectively. At $\log M \sim 1.5$ the same symbols illustrate the radii $R_{10}$ and $R_{20}$ obtained with $\beta = 1.1$. The thin lines show the relations from a hydrodynamic model (dashed-dotted; Langer 1989a) and hydrostatic interior models (long dashed; Schaerer & Maeder 1992)

Sofar all models have used plane parallel atmospheres for the outer boundary, which does not allow to take the strong mass outflow into account. Given our treatment of the outflowing envelope we can now also derive its radial extension. Since commonly radii of WR stars are defined at a Rosseland optical depth of typically 10 to 20

LTE atmospheric models, we have plotted the radius corresponding to these depths (Fig. 10). However, as a word of caution we remind the reader that direct comparisons with radii derived from non–LTE analysis may still be hampered if the hydrodynamic structures (particularly at large depths) adopted in both models differ significantly. Note that the velocity at $\tau = 20$ (10) is between 30 and 470 km s$^{-1}$ (100 and 920 km s$^{-1}$) for the interval plotted in Fig. 10. The hydrostatic layers are reached at optical depths of $\tau(R_\star) \sim 20$ to 130. Figure 10 shows that for WNE stars the subphotospheric layers (i.e. $\tau > 10$ or 20) may reach up to $\sim 2$ times the radius of the hydrostatic part, which corresponds to a temperature decrease of $\lesssim$ 0.15 dex, with respect to temperatures of $\log T_\star \sim 5.1$ – 5.15. The predicted decrease of the spherical extension with mass is simply due to the adopted mass-dependent mass loss rate, which implies a decrease of the wind density with decreasing mass, i.e. luminosity.

The definition of $R_\tau$ obviously depends on the envelope parameters ($\beta$, $v_\infty$, $\kappa$), but also on $\dot M$, which is however not treated as a free parameters (Sect. 2.3). As mentioned earlier, we note that the observations show a good agreement with the adopted mass loss rates (Hamann 1994), or even point to a somewhat higher values for $\dot M$. As we can see (Fig. 10) a lower $\beta$ yields slightly smaller radii. The terminal velocities, according to Eq. 2.3, are of the order of 2550–2680 km s$^{-1}$, which is $\sim 15$ % larger than typical observed values of WNE stars. Most important is clearly the wind opacity, for which we have taken the electron scattering opacity only. This should represent the minimum value for frequency independent continuum opacity. Provided the hydrodynamic structure of WR envelopes does not differ significantly from our description more realistic radii for WNE stars are thus quite probably underestimated by the mass-radius relation presented here.

The properties of the hydrostatic interior deserve one additional comment. Therefore we have also plotted the radius $R_\star$ of the hydrostatic interior. As expected this agrees well with previous calculations, although we now take the backwarming effect of the expanding envelope into account. This means that contrarily to WNL stars at $T_\star \lesssim 44000$ K, for WNE stars (the same holds for WC/WO stars, see Sect. 6) the outer boundary conditions have *no influence* on the interior structure, and in particular on its extension. This result is robust to any envelope parameter change, since all solutions rapidly converge in the radiative layers – the region with opacity peak is now located in the expanding envelope (cf. Sect. 4).

## 6. WC stars

In continuation from WNE stars discussed above, we now turn to WC stars. Both the wind velocities of WC stars, and their mass loss rates differ from the WNE phases. While the latter are a factor 5/3 larger for the same mass

Eq. 9 are smaller during most of the WC/WO phase than their WNE counterpart. Thus the winds are denser and the extension is increased.

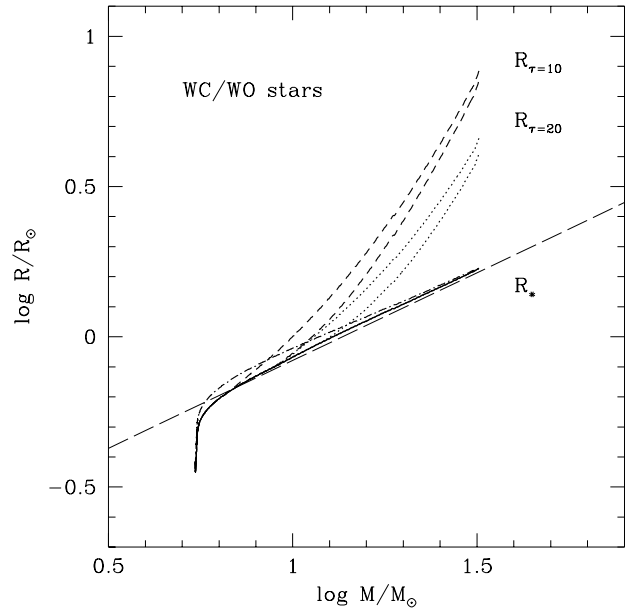

**Fig. 11.** Mass–radius relation for WC/WO stars including the spherical mass outflow. Same symbols as in Fig. 10. For each radius ($R_{10}$, $R_{20}$) a pair of curves illustrates a variation of $\beta$ from 2.1 (larger radius) to 1.1 (smaller $R$). Also shown is the relation for hydrostatic wind–free stars from Schaerer & Maeder (1992; thin long-dashed line). The dashed-dotted line illustrates the small radius ($R_\star$) increase due to CO opacities

This is illustrated in Fig. 11, where we show the mass–radius relation for WC/WO stars taking into account the spherically expanding envelope. The behaviour is as for WNE stars. We note that even with the low wind blanketing adopted, the radii corresponding to $\tau = 10$ are enlarged by up to a factor of 4 for the models with the largest mass loss rate. For this case the total optical depth of the expanding envelope amounts to $\tau \sim 160$–250 for $\beta = 1.1$–2.1. The main effect causing the decrease of spherical extension is due to the decrease of $\dot M \propto M^{2.5}$, while $v_\infty$ only increases by $\sim 0.4$ dex over the interval shown in Fig. 11. Also shown is the small influence of the parameter $\beta$. As for the WNE stars, we would like to point out that the radii obtained from this relation should provide a *lower limit* to the "core radii" derived from non–LTE analysis, since these are defined on the Rosseland optical depth scale by $\tau_{\rm Ross} \sim 10$–20, while our models only account for electron scattering opacity. The same words of caution (cf. Sect. 5) regarding the hydrodynamic structure of both non–LTE models and our approach are, however,

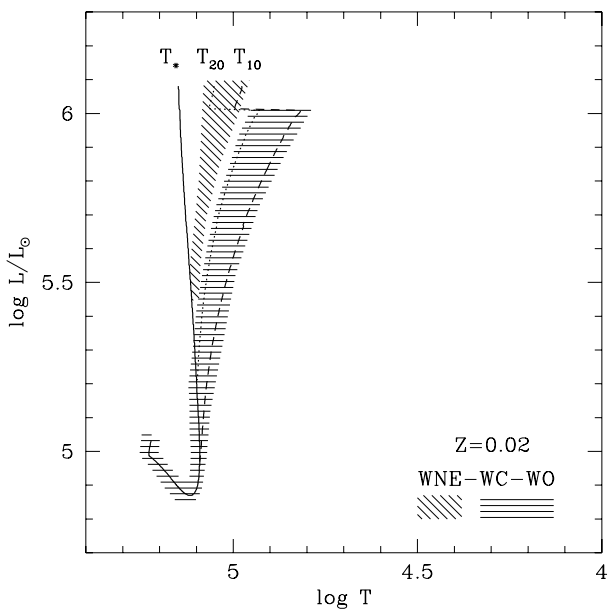
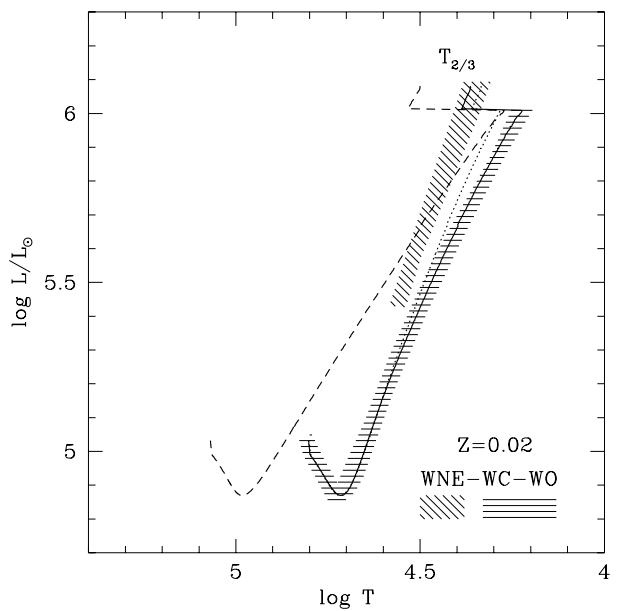

**Fig. 12.** HR–diagram for WNE and WC/WO stars including a spherically expanding envelope. The tracks are shown referring to different effective temperatures (cf. Sect. 2.1 for definitions) corresponding to the radius $R_\star$, and the subphotospheric radii $R_{\tau=20}$ and $R_{\tau=10}$ (same symbols as in Fig. 11). The shaded areas denote the WNE and WC/WO sequences respectively, which are obtained from different initial masses. The width of the sequences has been chosen to encompass the temperatures between $T_{20}$ and $T_{10}$

**Fig. 13.** HR–diagram for WNE and WC/WO stars including a spherically expanding envelope. Temperatures refer to an optical depth of $\tau = 2/3$. Solid line: preferred model adopting the Schaller et al. (1992) treatment including electron scattering and line opacity, and observed wind velocities (see Sect. 2). Dotted line: same as solid line but using $v_\infty = 2200$ km s$^{-1}$ following Schaller et al. Dashed line: $T_{2/3}$ obtained using observed values for $v_\infty$ but including only electron scattering. The shaded areas show the WNE and WC/WO sequence derived from the preferred model

also of concern here. Again, in the WC/WO phases, the interior structure is not modified by the presence of the outflowing envelope.

The HR–diagram covering the WNE and WC/WO phases of the 85 $M_\odot$ track is plotted in Figure 12. Shown are the different effective temperatures corresponding to the radius $R_\star$ and the subphotospheric radii $R_{\tau=20}$ and $R_{\tau=10}$ plotted in Fig. 11. Shaded areas indicate the position of the WNE and WC/WO sequences which are obtained for other initial masses. The width of these areas indicate the domain between $T_{20}$ and $T_{10}$.

As a comparison with previous work we also plot the HR–diagram referring to $T_{2/3}$ (Fig. 13). Accounting only for electron scattering yields the dashed line. The solid line shows the track obtained according to the treatment of Schaller et al. (1992) including electron scattering and line opacity in the wind, but using the observed wind velocities from Sect. 2. If instead we use a constant value of $v_\infty = 2200$ km s$^{-1}$ as in the original treatment of Schaller et al. we obtain the dotted line. Despite this difference the predictions for $T_{2/3}$, and in particular the slopes in the HR diagram agree very well. This can easily be explained by the relative importance of electron scattering and the lines to the total opacity, as well as their different dependence on $\dot M$ and $v_\infty$ (see Schaller et al. 1992). The shaded areas in Fig. 13 denote the WNE and WC/WO sequence as obtained from our preferred treatment (solid line).

### 6.1. Importance of CO opacities

In this context it is worth mentioning briefly the rôle of appropriate opacities for WC/WO stars. All evolutionary calculations for WR stars based on the recent OPAL data use opacities calculated with relative metal abundances corresponding to solar values. For evolved phases, in particular for WC stars, which show the products of He burning on their surface, this is clearly not valid anymore. We have therefore used the recent OPAL opacities for CO rich mixtures (Iglesias & Rogers 1993), which allow to correctly take the modified abundances of C and O into account. For typical WC model this results in a modest opacity enhancement ($\lesssim 20$ %) in the outer layers, which yields a radius increase of up to 10 % only for $R_\star$ (see Fig. 11). The He-burning lifetime of the 85 $M_\odot$ model is, however, increased by only 0.32 %. For most purposes standard opacities are therefore clearly sufficient.

Basic to our treatment is the assumption of radiative equilibrium in the expanding envelope. Although the same assumption is also made in non–LTE atmosphere models, which are used for comparisons in this work, one should be aware of a potential difficulty. Considering only the mechanical energy flux (neglecting the potential energy), we note that for the WN stars analysed by Hamann et al. (1993) this makes up between 0.4 % to 60 % of the radiative luminosity for the most extreme case. The influence of this assumption on the derived stellar parameters is presently unknown. With the revision of luminosities (upward) and mass loss rates (downward) suggested by Schmutz (1995) the contribution of the mechanical flux to the total luminosity is, however, drasticly reduced.

For the most luminous WNE and WC models from our evolutionary tracks, the kinetic energy flux may also reach up to $\sim 20$ % and 8 % of the radiative flux respectively. Accounting for the mechanical energy flux and also work done against the gravitational potential the luminosity should be decreased by an amount raging between $\sim 0.14$ (0.25) dex at $\log L/L_\odot = 6.$ and 0.02 (0.08) dex at $\log L/L_\odot = 5.5$ referring to optical depths of $\tau = 10$ ($\tau = 2/3$) respectively. As expected, differences are only of some importance at the highest luminosities, corresponding to $\dot{M} \gtrsim 10^{-4} M_\odot \, \mathrm{yr}^{-1}$.

## 7. Comparison with observations

A detailed comparison between predictions from evolutionary models and observations of WR stars, including luminosities, wind corrected temperatures ($T_{2/3}$), abundances, and population statistics has recently been presented by Maeder & Meynet (1994). As shown above, the inclusion of the expanding envelope yields, for all cases except maybe WNL stars (cf. Sect. 4), the same results as the models used by Maeder & Meynet. Thus their conclusions also apply to our models. On the other hand basically two new predictions are made in this work. These will briefly be compared to observations in this paragraph.

### 7.1. WN stars

In Sect. 4 we have suggested the existence of a region located at $4.4 \lesssim \log T_\star \lesssim 4.65$, where WNL models are temporarily concentrated before evolving to higher temperatures. Interestingly the strip suggested by the models with strong backwarming is found to be at the same "core" temperatures, where most observed WN stars still having hydrogen are placed by non–LTE spectral analysis (Hamann et al. 1993, Crowther et al. 1995b). Whether this is a pure coincidence is not clear. Statistical arguments about the relative WN population at different temperatures are surely not yet possible. On the other hand theoretically there is no relation between the value of the H surface abundance and the proposed "WNL strip", which show hydrogen on their surface (cf. Fig. 9). The location of the strip is only determined by the temperature. It is however clear, that the motion towards larger temperatures is related with an increasing (decreasing) He (H) abundance. Therefore stars located in the strip, and even more redwards of it, always have a non-zero H abundance. This seems to be in disagreement with some WNE-w stars analysed by Hamann et al. (1993).

Regarding the extension of WNL stars, except WN+abs subtypes, we note that Crowther et al. (1995b) find values of $R_{2/3}/R_\star \sim 1.2$–2.1 in contrast with $R_{2/3}/R_\star \lesssim 1.1$ from Hamann et al. (1993), who claim that the continuum emerges from nearly hydrostatic layers. Since both authors, however, use the same wind velocity law with $\beta = 1$, this indicates that even for WNL stars the definition of the core radius depends on the adopted structure in the low velocity region, which is hardly constrained by the observations. For WN stars with strong lines the ambiguous assignment of core radii from non–LTE models was already shown by Hillier (1991; see also Schmutz et al. 1992). This must be taken into account when comparing core radii and temperatures.

### 7.2. WC stars

As a second point our new calculations predict, for the first time, what could be called sub-photospheric radii, i.e. wavelength independent radii for WR stars with strong mass outflows. These radii, as well as the associated effective temperatures, should allow a comparison to the so-called core radii determined from non–LTE spectral analyses of observed WC stars. As mentioned above, it has, however, to be kept in mind that even for non–LTE analyses the definition of a core radius may not be unique (Hillier 1991, Schmutz et al. 1992).

We compare our results with the analysis of 25 Galactic WC stars of Koesterke & Hamann (1994). In their method two parameters are determined from line fits, taken apart the abundances. The basic parameters are the "transformed radius"

$$R_t = R_\star \left[ \frac{v_\infty}{2500 \, \mathrm{km \, s^{-1}}} \frac{10^{-4} M_\odot \, \mathrm{yr}^{-1}}{\dot{M}} \right]^{2/3}, \quad (12)$$

which is a measure of the wind density, and the effective temperature $T_\star$. The corresponding "core" radius $R_\star$ refers to depths of $\tau_\mathrm{Ross} \sim 10$–20. To derive the other physical quantities ($R_\star$, $L_\star$, $\dot{M}$) requires the knowledge of the absolute visual magnitude, and the terminal velocity. The most direct comparison with their results is thus made in the $T_\star$–$R_t$ plane.

The evolutionary track of the 85 $M_\odot$ model in the $T_\star$–$R_t$ plane is shown in Fig. 14. We have plotted the track referring to both the subphotospheric radii $R_{10}$ and $R_{20}$, but also using the radius of the wind–free hydrostatic model. For a comparison we have added a track where the wind

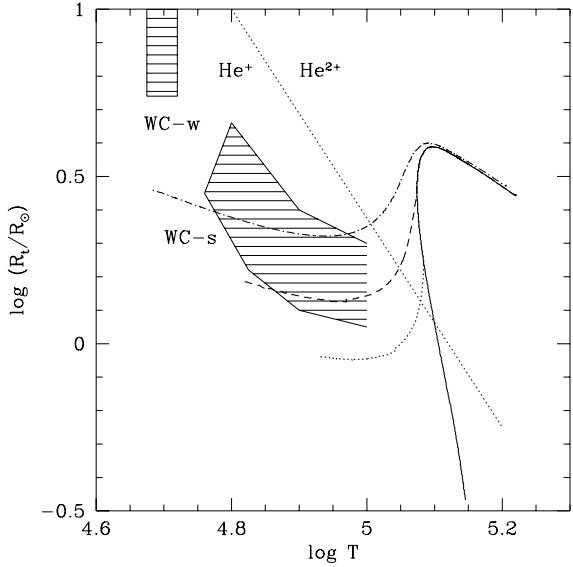

**Fig. 14.** Evolutionary track for WC/WO stars in the $T$–$R_t$ plane. As in Figs. 11 and ?? the effective temperatures and radii refer to $\tau = 10$ (short-dashed), 20 (dotted), and the wind–free hydrostatic interior (solid). The dashed-dotted line is obtained with a wind opacity of $2\sigma_e$ for $\tau = 10$. The WC-w and WC-s groups from the non–LTE analyses of Koesterke & Hamann (1994) are indicated by the shaded areas. Regions where He is completely ionised or recombines to He$^+$ are separated by the thin dotted line (cf. Schmutz et al. 1992)

opacity is $2\sigma_e$. The evolution proceeds essentially from the left to the right (increasing $T$). Note that independently of their initial mass and metallicity all WC models populate the same line in $T$–$R_t$ plane, since these quantities are mainly determined by the mass loss rate, $L_\star$, and $R_\star$, which are all related to the stellar mass[3] (cf. Schaerer & Maeder 1992). Depending on their initial mass and metallicity the starting position of the WC phase on this line, and in particular the surface abundances, do however vary considerably.

Shaded areas indicate the domains of the analysed WC 5 to WC 7 stars from Koesterke & Hamann (1994), for which they obtain good fits. Obviously the WC-w stars showing low wind densities and cooler effective temperatures are not matched by our models. If their continua should indeed arise from almost hydrostatic layers, as claimed by Koesterke & Hamann, the difference with our results cannot be due to an underestimate of the wind opacity, which would otherwise lead to lower temperatures and larger radii.

On the other hand we note that our predictions referring to $\tau = 10$ partly cover the same region as the WC 5 to 7 subtypes with strong lines (WC-s). Subtypes ear-higher temperatures. With ongoing evolution (C+O)/He increases, which implies that on one particular track later WC subtypes (cf. Smith & Maeder 1991) should be located at lower temperatures. Surprisingly Koesterke & Hamann indicate the opposite trend for WC 5 and 6 subtypes. As explained above, models with various metallicities and initial masses show, however, important surface abundance variations in the $T$–$R_t$ diagram. Therefore one theoretically expects to find different WC subtypes mixed up at the same position in this diagram.

---
[3] The obvious assumption is that the wind parameter $\beta$ and the opacity $\kappa$ are unchanged.

In short we find that our evolutionary models, taking into account the strong mass outflow, roughly reproduces the wind densities and effective temperatures of observed WC-s stars. Several points, however, show still differences with the results from non–LTE spectral analyses. In particular the existence of WC stars with "weak winds" (WC-w) at relatively low temperatures ($T_\star \sim 50$ kK) cannot be understood with present day evolutionary models. Particularly striking is also the large spread of carbon abundances derived by Koesterke & Hamann (1994) for a given WC subtype, contrarily to several previous determinations (cf. Smith & Hummer 1988, Eenens & Williams 1992, Smith & Maeder 1991), which have also led to a simple explanation of the observed subtype gradient (Smith & Maeder). Whether this may be related to the above-mentioned difficulties in comparing non–LTE model analysis with evolutionary model predictions is, however, not clear yet.

## 8. Conclusions

A simple analytic method to describe the structure of a spherically expanding envelope with strong mass outflow has been presented in this work. We solve for radiative equilibrium in the grey case and for mass conservation, while the momentum equation is replaced by adopting a parametrised velocity law. The structure is consistently connected to the hydrostatic stellar interior and allows us to study the influence of modified boundary conditions ("backwarming") on the stellar structure and evolution (Sect. 2). On the same ground our treatment also permits to determine subphotospheric "core" radii and associated effective temperatures, which characterise the entire emergent spectrum of WR stars.

We have applied the new method to evolutionary models of WR stars. For WR stars showing still hydrogen on their surface (WNL's), models with strong backwarming predict a transient concentration in a strip located at temperatures $25000 \lesssim T_\star \lesssim 45000$ K, indepently of the luminosity. This is due to a small convective zone close to the surface, which is influenced by the outer boundary conditions and maintains a low density gradient, thereby allowing to temporarily stabilise the stellar radius on the blueward track (see Sect. 4). Surprisingly this strip coincides quite well with the "degenerate " location, where hand, if a moderate backwarming is adopted, the tracks in the HR–diagram are basically unchanged with respect to the usual treatment using plane parallel atmospheres, and the "WNL strip" disappears.

In any case, including all the WR phases up to the end of He-burning, the interior structure and hence the evolution is never modified by the presence of an outflowing envelope (see Sects. 4 to 6). It is well known that the dominant process determining the evolution in this domain is mass loss (cf. Chiosi & Maeder 1986). We have now shown, that it is indeed sufficient to consider only the effect of a progressive reduction of the stellar mass. The other consequences of mass loss, including backwarming from a spherically expanding envelope, have *no effect* on the deep interior structure. This does, however, not generally hold for the stellar parameters (radii, effective temperatures) and in particular does not apply to the spectral appearance, which obviously strongly depends on the "details" of the mass loss phenomenon.

For hydrogen free (WNE, WC and WO) stars with strong mass loss we have calculated subphotospheric radii corresponding to optical depths of $\tau \sim 10$–$20$ (Sects. 5 and 6). The radius increase with respect to wind-free stars (both hydrostatic or hydrodynamic models) may be up to a factor of $\sim 4$ for the most luminous WNE or WC stars. Provided the hydrodynamic structures (particularly at large depths) adopted in both our approach and in non–LTE models do not differ significantly, our subphotospheric radii should be comparable with "core" radii derived from non–LTE spectral analyses. Comparisons with observations show a better agreement than previous models (Sect. 7). Future progress awaits full hydrodynamic calculations of coupled interior and atmosphere models.

Subphotospheric radii also allow a direct assignment of spectra to evolutionary tracks. Compared with wind-free evolutionary models this implies e.g. H (He) ionising fluxes (derived from the non–LTE models of Schmutz et al. 1992), which are reduced by up to $\sim 30$ (60) %. Finally our treatment of the envelope now permits to pursue the study of the spectral and interior evolution of massive stars presented by Schaerer (1995) and Schaerer et al. (1995ab) to post-MS phases including WN, WC and WO stars.

Last, but not least, we have pointed out the possible importance of the iron opacity peak for the acceleration of WR winds in the optically thick part (Sect. 4). This may well be essential for the understanding of the dynamics of Wolf–Rayet winds.

*Acknowledgements.* I am grateful to Dr. Georges Meynet for stimulating discussions and careful reading of the manuscript. Particular thanks are addressed to Prof. André Maeder for his continuous encouragement and support. Financial assistance from the Swiss National Found of Scientific Research is also acknowledged.